\documentclass[%
 aip,
 amsmath,amssymb,
 reprint,%
floatfix
]{revtex4-1}

\pdfoutput=1

\usepackage{graphicx}% Include figure files
\usepackage{dcolumn}% Align table columns on decimal point
\usepackage{bm}% bold math
\usepackage[utf8]{inputenc}
\usepackage[T1]{fontenc}
\usepackage{mathptmx}
\usepackage{etoolbox}

\usepackage{placeins}
\usepackage{float}
\usepackage{afterpage}
\usepackage{capt-of}
\makeatletter
\def\@email#1#2{%
 \endgroup
 \patchcmd{\titleblock@produce}
  {\frontmatter@RRAPformat}
  {\frontmatter@RRAPformat{\produce@RRAP{*#1\href{mailto:#2}{#2}}}\frontmatter@RRAPformat}
  {}{}
}%
\makeatother

\begin{document}

\title{Design and construction of a quantum matter synthesizer}
\author{Jonathan Trisnadi}
\affiliation{James Franck Institute, Enrico Fermi Institute, and Department of Physics, \\ The University of Chicago, Chicago, IL 60637, USA}
\author{Mingjiamei Zhang}
\affiliation{James Franck Institute, Enrico Fermi Institute, and Department of Physics, \\ The University of Chicago, Chicago, IL 60637, USA}
\author{Lauren Weiss}
\affiliation{James Franck Institute, Enrico Fermi Institute, and Department of Physics, \\ The University of Chicago, Chicago, IL 60637, USA}
\author{Cheng Chin}
\email{cchin@uchicago.edu}
\affiliation{James Franck Institute, Enrico Fermi Institute, and Department of Physics, \\ The University of Chicago, Chicago, IL 60637, USA}

\date{\today}

\begin{abstract}
The quantum matter synthesizer (QMS) is a new quantum simulation platform in which individual particles in a lattice can be resolved and re-arranged into arbitrary patterns.
The ability to spatially manipulate ultracold atoms and control their tunneling and interactions at the single-particle level allows full control of a many-body quantum system.
We present the design and characterization of the QMS, which integrates into a single ultra-stable apparatus a two-dimensional optical lattice, a moving optical tweezer array formed by a digital micromirror device, and site-resolved atomic imaging.
We demonstrate excellent mechanical stability between the lattice and tweezer array with relative fluctuations below 10~nm, high-speed real-time control of the tweezer array at a 2.52~kHz refresh rate, and diffraction-limited imaging at a resolution of 655~nm.
The QMS also features new technologies and schemes such as nanotextured anti-reflective windows and all-optical long-distance transport of atoms.
\end{abstract}

\maketitle

\section{Introduction}
Over the last two decades, quantum simulation using ultracold atoms has proven to be a fruitful endeavor toward the exploration of many quantum phenomena.
Progress in this field has accelerated tremendously thanks to the growing experimental toolkit that enables precise control of the many-body system and access to a rich variety of research topics~\cite{altman2021quantum, schafer2020tools}.

%This recent trend has been bolstered by new and exciting technologies that push the boundaries in terms of the variety of physical systems accessible, as well as the fidelity in the initial state preparation.

Two mainstays of cold atom quantum simulation platforms are optical lattices~\cite{esslinger2010fermi,  gross2017quantum} and moving optical tweezer arrays~\cite{kaufman2021quantum}.
Experiments with atoms trapped in optical lattices have demonstrated phenomena such as quantum phase transitions~\cite{greiner2002quantum, gemelke2009situ, bakr2010probing, guardado2018probing}, quantum transport~\cite{brantut2012conduction, preiss2015strongly}, and artificial gauge fields~\cite{aidelsburger2013realization, zohar2015quantum, cooper2019topological}.
%Double check I Bloch paper and prefer to cite those examples
A powerful feature of optical lattice systems is the tunability of the tunneling and interaction energy, which allows tailoring of the Hamiltonian of atoms~\cite{gross2017quantum}.
Novel lattice geometries can be realized with appropriate configuration of the laser beam interference.
%~\cite{windpassinger2013engineering, becker2010ultracold, yamamoto2020single, yang2021site}.
The integration of high-resolution imaging further enables the detection of atom occupancy in the lattice at single-site level~\cite{bakr2009quantum, sherson2010single}.
%, cheuk2015quantum, haller2015single, parsons2015site, miranda2015site, edge2015imaging, yamamoto2016ytterbium, kwon2022site}.
%The connectivity of the lattice can play an important role.
Moving tweezer arrays formed by deflecting and focusing multiple laser beams have proven useful for initiating atom distributions with reliable filling~\cite{lester2015rapid, barredo2016atom, endres2016atom, lee2017defect}, realizing quantum information processing~\cite{omran2019generation,bluvstein2022quantum}, and assembling molecules from single atoms~\cite{PhysRevX.9.021039}.
%Bernien2017probing is more about quantum spin models- could remove
%omran2019generation is about large scale entanglement between Rydberg atoms and superconducting qubits
% Rydberg array citation section- omran2019generation, 
%By combining optical lattices with optical tweezer arrays, excellent single-particle control has been demonstrated\cite{kaufman2022}. % need to fix this citation?

%Yet so far, these two technologies have remained relatively isolated from each other. 
%Typically, quantum gas microscopes employ conventional optical lattices with high filling fractions afforded by quantum degeneracy, and any desired defects are produced by “poking holes” with a blue-detuned laser. 
%The downside here being that the prepared initial states are constrained by the initial atom distribution and cannot have arbitrary occupancy at each site. 
%AOD tweezer-array experiments do not suffer from this constraint, but must be operated at a large $>1$~$\mu$m spacing in order to avoid resonant heating at the trap frequency, which limits its application in studying tunneling dynamics in Hubbard-regime lattices.

Here we report on the construction of a quantum matter synthesizer (QMS), a quantum simulation apparatus that combines atoms in an optical lattice with the dynamic rearrangement capability of tweezer arrays. 
The marriage of these two technologies allows arbitrary initial state preparation using tweezers while maintaining the uniform and closely-spaced optical lattice site potentials for Hubbard model studies~\cite{kaufman2022}. 
% Furthermore, the experimental cycle time can be made $\sim$~2-10 times shorter by bypassing the long evaporation time necessary in quantum degenerate experiments.
%We detail the key components in the construction of the QMS.
In this paper, we present the design, construction, and characterization of the QMS, with a focus on the mechanical stability of the optical potentials, the cooling and site-resolved imaging of atoms, and the generation of moving tweezers based on a digital micromirror device (DMD).

In the following sections, we describe the key components of the QMS, starting with the mechanical design in Sec.~\ref{secQMS}.
The precise control of an atom in a lattice site requires exquisite stability between the lattice and tweezer potentials. 
We gain the required stability by implementing a rigid mechanical structure and sharing the same optical path for both potentials, see Sec.~\ref{secMech}. 
Next, the triangular lattice and optical tweezer potentials are prepared to confine and pattern the atoms in a nanotextured glass cell, described in Sec.~\ref{secGlasscell}.
The lattice beams are described in Sec.~\ref{secLattice} and the DMD potential in Sec.~\ref{secDMD}.
Finally, we will discuss atom transport and \textit{in situ} imaging of atoms in the lattices in Sec.~\ref{secTran} and Sec.~\ref{secImg}, respectively.

% In the Sec.II we describe the idea of the quantum matter synthesizer and its basic design.
% Next, we detail the design and construction of a custom stainless steel optics structure that provides excellent mechanical stability.
% In Sec.IV we provides an overview of our special nano-textured glass science cell.
% In Sec.V we describe our 2D triangular lattice setup and its characteristics.
% The digital micromirror device is discussed in the sixth section, including its high spatial and temporal resolution.
% In the seventh section we discuss our implementation of an all-optical long-distance 28~cm transport scheme for the atom cloud using a moving 1D lattice.
% Lastly, we describe our cooling and site-resolved imaging process in Section 8.

\section{Working principle of the quantum matter synthesizer}
\label{secQMS}

\begin{figure*}[htp]
\centering
\includegraphics[width=150mm]{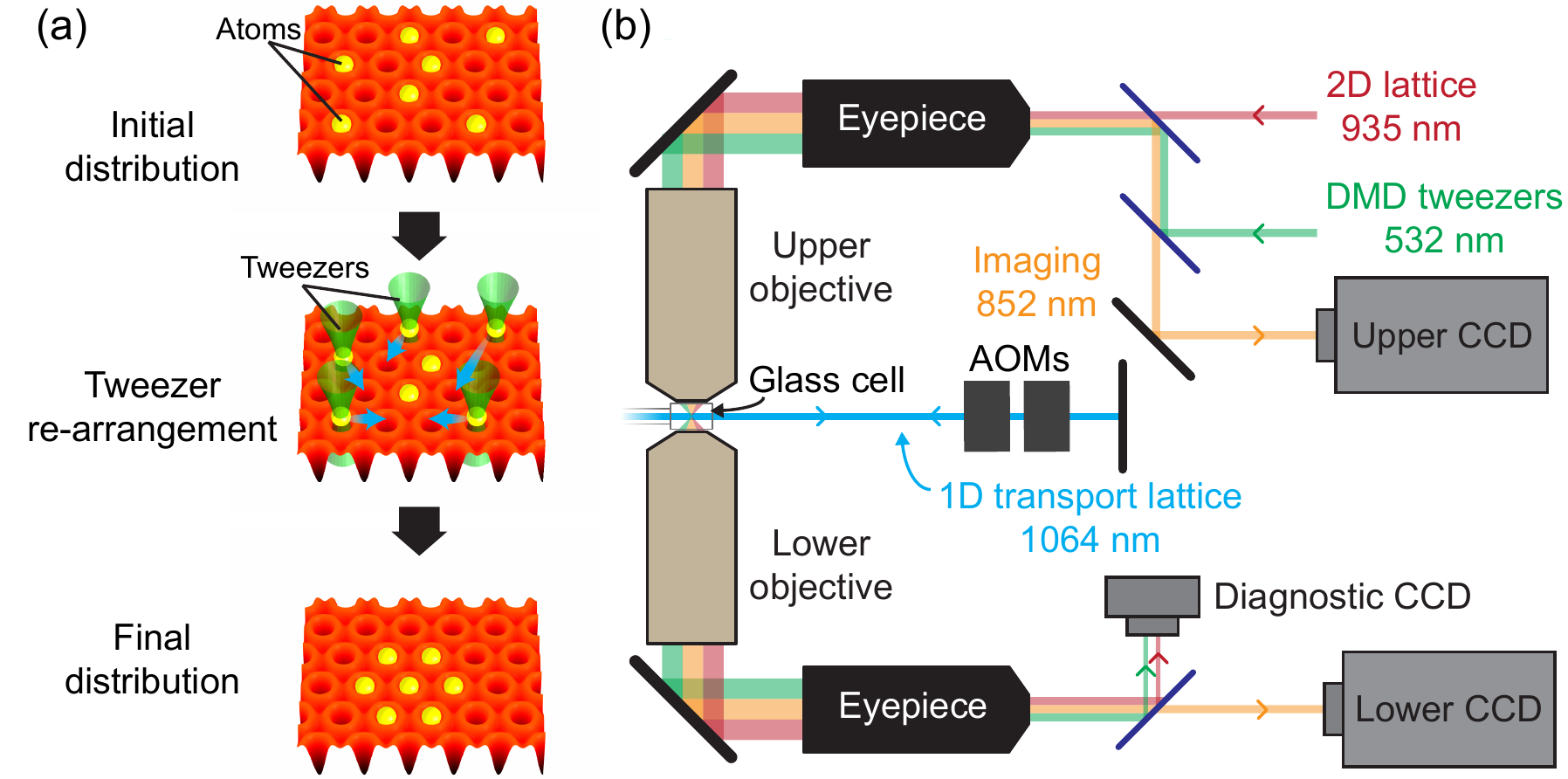}
\caption{Working principle of the quantum matter synthesizer and general experimental design.
(a) Within a single experiment cycle, the QMS images the initial atomic distribution in a 2D-lattice and generates a set of moves for a controllable tweezer array which re-arranges the atoms into a pre-specified final configuration.
(b) The general optics scheme for the system is shown, featuring two high-numerical-aperture microscope objectives and three CCD cameras.
The 2D lattice beams and the tweezer light are sent through the top objective, which also collects fluorescence from the atoms to image them on the upper CCD.
The lower objective sends the trapping light to a diagnostic CCD, which images the optical potentials in order to diagnose aberrations and positional drift.
The emitted photons from atoms are also collected on the lower CCD which serves as a secondary imaging camera.
}
\label{figQMS}
\end{figure*}

The QMS operates by assembling and patterning the sample via stereotactic manipulation of individual atoms.
This is realized through highly coordinated and stable control of optical tweezers in combination with high resolution imaging of single atoms in an optical lattice. 
This capability will greatly increase the flexibility and fidelity of quantum simulation and allow for fast and efficient removal of entropy to prepare atoms in the quantum degenerate regime\cite{weiss2004another}.
Moreover, in contrast to traditional cold atom experiments which typically study ground state or thermal equilibrium distributions of atoms, the QMS can tailor far-from-equilibrium, designer quantum states. 

%The experimental apparatus is sketched in Fig.~\ref{figQMS}b. 

The operation of our QMS consists of three steps:

\noindent \textbf{Step 1: Sample preparation.}
The QMS starts by collecting precooled Cs atoms in a magneto-optical trap (MOT), cooling them with degenerate Raman sideband cooling (dRSC), and then transferring them into a glass science cell using a moving 1D optical lattice.
In the glass cell, the atoms are loaded into a 2D optical lattice, where each site has random occupancy.

\noindent \textbf{Step 2: Ground state cooling and imaging.}
We first determine the atomic distribution in a deep 2D lattice based on site-resolving \textit{in situ} imaging.
During the imaging, the dRSC is employed to suppress tunneling and heating of the atoms.
The photons emitted during cooling cycles are collected to image the site occupancies in the lattice (see Section~\ref{secImg}).
With an imaging resolution of 655~nm, small compared to the lattice constant 881~nm, a high fidelity single site resolution can be reached.

\noindent \textbf{Step 3: Rearranging atoms.}
Based on the measured distribution in the lattices, we rearrange atoms into a desired pattern.
To move the atoms to new locations, we use an array of optical tweezers at 532~nm prepared by a DMD.
We first tune up the tweezer potential slowly to securely localize the atoms in the tweezer potential.
We then drag atoms to new lattice sites adiabatically by video streaming a series of light patterns from the DMD.
Throughout the relocation process, atoms are tightly confined in the tweezers to prevent tunneling. 
At the end of the process, atoms are released back to the lattice sites and cooled back to the ground state to remove excess entropy gained during the motion (see Fig.~\ref{figQMS}a).

\section{Mechanical structure and stability}
\label{secMech}

\begin{figure*}
\centering
\includegraphics[width=155mm]{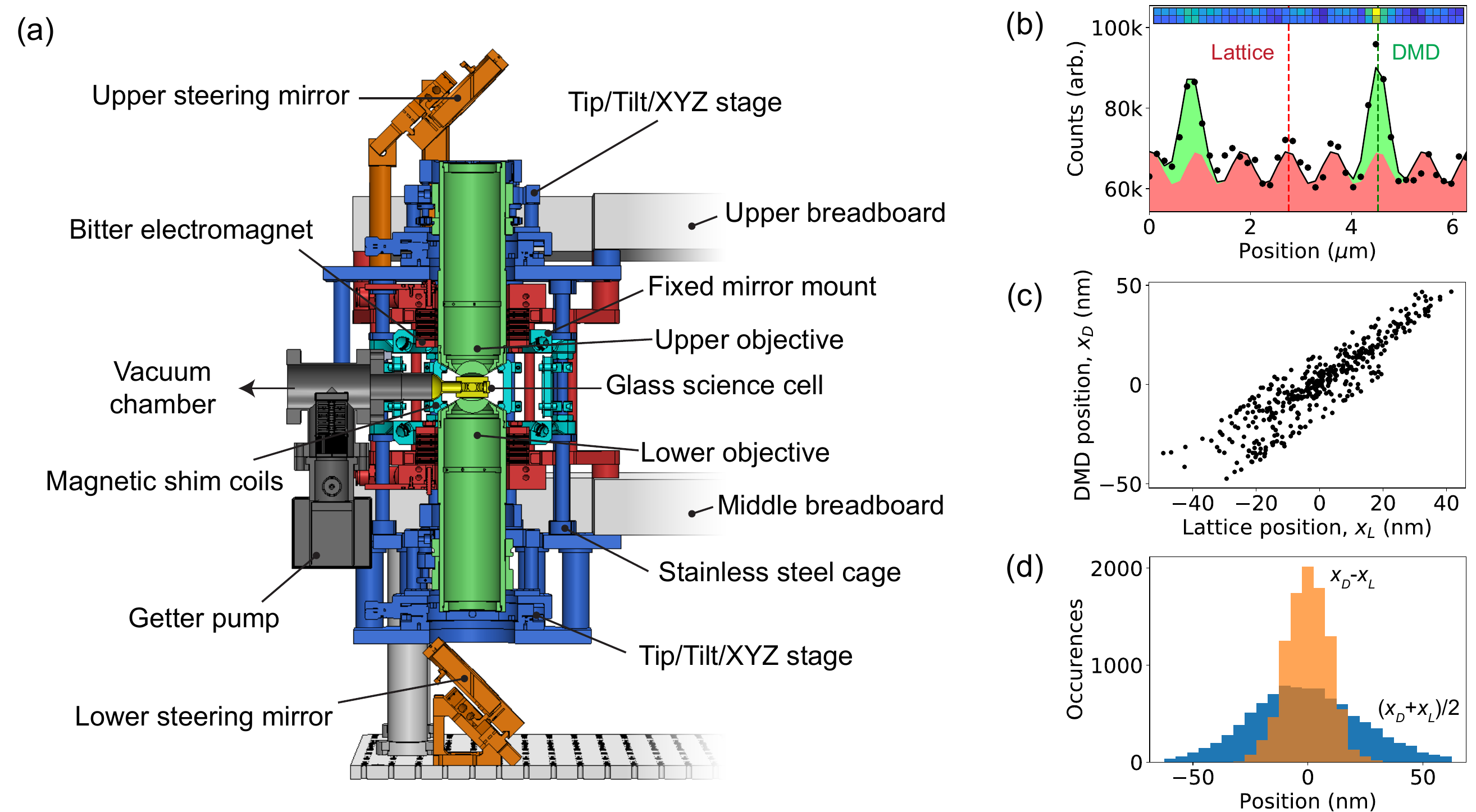}
\caption{Mechanical design and stability of microscope optical system.
(a) A thick stainless steel cage (blue) holds the two microscope objectives (green) around the glass cell (yellow). Nearby optics and components are also shown, including 45$^\circ$ steering mirrors (orange), Bitter electromagnets and mounts (red), fixed mirror mounts and magnetic shim coil (cyan), the vacuum chamber (dark grey), and the 3-layer breadboard structure (light grey).
(b) We analyze the mechanical stability between the tweezer light and the lattice light by imaging both simultaneously on the diagnostic CCD and observing their positions over time. We fit the lattice position (red dashed line) and the tweezer position (green dashed line) using the CCD image (inset).
(c) The deviations of the tweezer position $x_D$ and lattice position $x_L$ are highly correlated as a result of their shared beam paths. Data are shown for a measurement that is 1~second in duration.
(d) The difference in lattice and tweezer positions $x_D-x_L$ (orange) shows a much tighter distribution than the mean position of the lattice and DMD (blue). The RMS noise of $x_D-x_L$ is 9~nm, which is much smaller than their average fluctuation of 26~nm. Data is taken over a 20~second long measurement.
}
\label{figMechanical}
\end{figure*}

%\textit{Short intro paragraph about imaging and addressing requiring ~1 micron stability, and how our design emphasizes common optics between lattice and tweezers (and we measure using high-resolution imaging).}
In order to image, trap, and manipulate atoms at the site-resolved level, the relative mechanical stability between the optical lattice at 935 nm, the tweezer array at 532 nm, and the atom fluorescence at 852 nm must be kept within the spatial extent of the on-site wavefunction in our system (approx. 30~nm). 
%a 10~nm scale (approximate size of the atomic wavefunction).
%We design and construct an ultra-stable optical setup in which
To this end, we prepare the imaging, lattice, and tweezer beam paths such that all pass through the microscope optics in order to reduce their relative mechanical instability (see Fig.~\ref{figQMS}b).
%mention putting beams into objectives -- 10 nm, refer to Fig 1b

The optical system is comprised of two identical microscopes placed symmetrically above and below the atom sample.
The objectives (Special Optics, Inc.) are custom-designed to offer diffraction-limited performance at all the relevant wavelengths of 532~nm, 852~nm, and 935~nm.
The objectives have a numerical aperture of $\mathrm{NA}=0.8$ and a working distance of 1.05~cm.
%The working distances are $\approx$1~cm, which is long compared to most microscope experiments that utilize solid-immersion lenses, so the technical complexity of trapping atoms very close to an optical surface can be avoided.
This dual microscope configuration allows us to image the tweezer and lattice light using the diagnostic CCD in order to analyze the quality of the optical potentials on the atoms.
%An advantage of having a dual microscope is the ability to easily image the tweezer and lattice light using the diagnostic CCD in order to analyze the quality of the optical potential on the atoms.
%perform fast and reliable analysis on the quality of the projected optical potentials by directly taking images of the light projected through both objectives (e.g. by imaging the light on the lower diagnostic CCD).
%This is in contrast to most experiments which use a single objective and rely on the atomic response to the light field in order to diagnose optical aberrations.

%High-fidelity atom re-arrangement requires that the relative instability between the optical tweezer array and the optical lattice should be as small as possible.
%We target a passive mechanical stability at the 10~nm level, which is smaller than the size of a ground state atomic wave function and much smaller than the 890~nm lattice constant.
%We design and construct 
An ultra-stable stainless steel cage around the glass cell holds the two microscope objectives and other supporting components (Fig.~\ref{figMechanical}a).
By connecting the two objectives via a cage, the relative vibrational noise between these two sensitive optics is greatly reduced.
%We use a three-level-breadboard structure to provide a solid foundation to mount the cage at various heights.
%First, the objectives are each mounted on tip/tilt/X/Y/Z stages to provide all the necessary degrees of freedom for fine alignment.
%These stages are then mounted to a stainless steel plate.
%The two plates are connected to each other through four 0.625” diameter stainless steel posts and eight 0.5” diameter posts.
%Finally, the plates are mounted to the upper- and middle-layer breadboards.
In addition to the objectives, the stainless steel cage also supports auxiliary mirror mounts.
A nearby water-cooled Bitter electromagnet is mounted on a separate structure to avoid acoustic noise caused by the water flow or large magnetic field quenches\cite{sabulsky2013efficient}.
%See the caption of Fig.~2a for a complete list of components.

We test the relative mechanical stability of the 935~nm optical lattice and the 532~nm tweezers by imaging them on the lower diagnostic CCD at a fast frame rate of 650~Hz.
%We do so by turning on the 935~nm lattice and two 532~nm tweezers, and image them all on the lower diagnostic CCD at a high 650~Hz frame rate.
By fitting the recorded images of the lattice sites and the tweezers, we track the variations of their positions $x_L$ and $x_D$ over time with high precision (see Fig.~\ref{figMechanical}b).
Details about the optical setup of the imaging, the lattice, and the tweezers can be found in Sections~\ref{secLattice} and \ref{secDMD}.

We observe that the two optical potentials experience highly correlated motion, indicating the common-mode behavior of the optical paths (see Fig.~\ref{figMechanical}c).
%We take a dataset over 20~seconds long and look at the RMS behavior in 1~second long windows.
While $x_L$ and $x_D$ display a root-mean-square instability of 26~nm, their relative instability is only 9~nm (Fig.~\ref{figMechanical}d), smaller than the expected on-site wavefunction extent.
Thus, the small relative instability makes the QMS amenable to reliable arrangement of atoms in the lattice using tweezers.
%We find that while the absolute position of each light source reaches an RMS value of 26~nm, the relative RMS noise is much smaller at only 9~nm (Fig.~2d), which is smaller than the atomic wavefunction size and much smaller than the lattice constant.

\section{Nanotextured glass cell}
\label{secGlasscell}

%The high-numerical aperture objectives transmit and receive light to and from the focal point at large angles. 
%In order to maximize the transmission and collection efficiency of our high-N.A. objectives, the glass science cell is constructed with special nanotextured windows that can provide good broadband anti-reflectance at large angles of incidence.
In order to maximize the optical transmission at wide acceptance angles for imaging, projection, and lattice formation at different wavelengths, we adopt a glass science cell (Precision Glassblowing Inc.) constructed with special nanotextured windows, rather than a traditional polymer-coated solution.
%In order to maximize the trap depth and photon collection efficiency, we use a glass science cell with nano-textured windows that provides good broadband anti-reflectance at large angles of incidence.
The windows (TelAztec LLC) provide excellent broadband anti-reflection (AR) in the 532~nm to 1064~nm range for angles of incidence up to 45$^\circ$.
%In particular, the AR properties must be broadband in order to accommodate the relevant wavelengths, which range from 532~nm to 1064~nm.
%Secondly, because of the high numerical aperture of the objectives, the windows must display good AR performance even at large 45$^\circ$ angles of incidence in order to maximize the photon collection efficiency.
%Typical polymer-coated solutions do not cover our required range of wavelengths and have reflectivities of $\sim 5\%$ when the incident angle exceeds $45^\circ$.

\begin{figure}
\centering
\includegraphics[width=85mm]{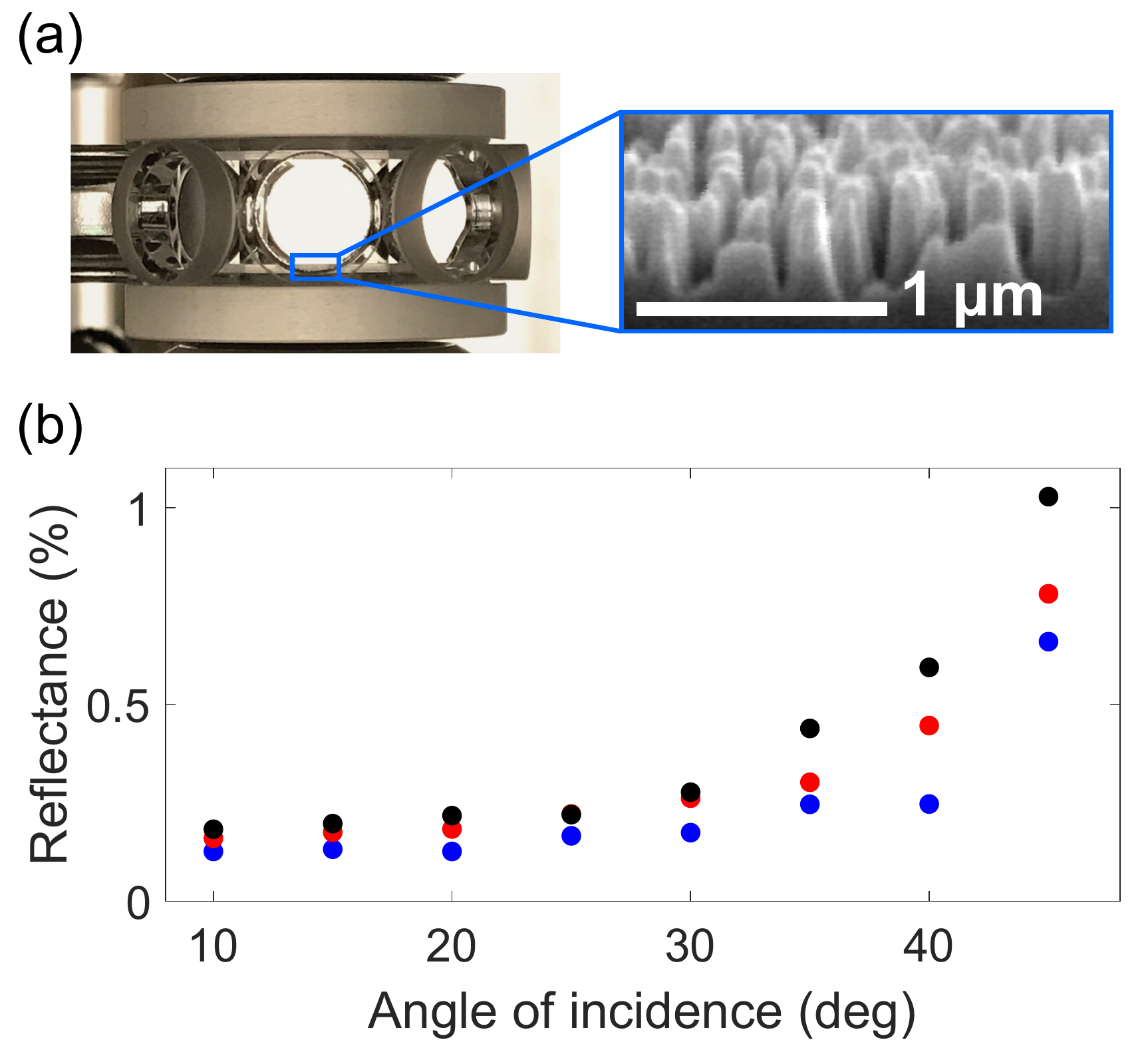}
\caption{Optical properties of nanotextured windows.
(a) Photo of the glass science cell (left) and a scanning electron microscope (SEM) image of the nanotextured surface (right). SEM photo courtesy of TelAztec LLC.
(b) The nanotextured surface gives rise to low broadband reflectance over a large range of angles of incidence. 
Shown are the percent reflected from both surfaces of a window for 700 (blue), 852 (black), and 935~nm (red).
}
\label{figGlasscell}
\end{figure}

%Rather than a traditional polymer-coated solution, we choose windows whose surfaces are treated by a nano-etching procedure that results in surface texturing with features in the $\sim100$~nm lengthscale (see Fig.~3a).
%The nanotexture consists of pillar-like structures on the scale of 100~nm.
The glass cell window is textured with roughly 100~nm protuberances that have a large degree of randomness in the size and spacing, which contribute to their broadband, wide-angle AR performance (Fig.~\ref{figGlasscell}a).
%Intuitively, 
The nanotexturing provides a smooth transition in the index of refraction from air to glass, thus avoiding the large mismatch that causes strong reflections.
These nanostructures are similar to the so-called ``motheye" metamaterials that typically consist of a 2D array of pillars which are uniform in size and work well for narrowband applications~\cite{wilson1982optical, hobbs2013contamination}.

We measure the window reflectivities (combined front- and back-reflections) at various wavelengths and verify that their reflectivity is $1\%$ or less at angles of incidence up to 45$^\circ$ (Fig.~\ref{figGlasscell}b).
Lastly,we also note that due to the absence of polymer coating on the windows, the glass cell can be baked to higher temperatures in excess of $200^\circ$~C than the windows with dielectric coating, which promises better vacuum.

\section{2D triangular lattice}
\label{secLattice}

%We use optical trapping to confine atoms in the glass cell.
Horizontal trapping of atoms in the glass cell is provided by a 2D triangular lattice formed by interfering three laser beams at wavelength $\lambda=$~935~nm.
%~\cite{becker2010ultracold}.
Here 935~nm is a so-called ``magic wavelength'' for the Cs D$_2$ transition at which atoms in the ground state and $6\mathrm{P}_{3/2}$ excited state experience the same lattice potential landscape~\cite{le2013dynamical}.
A 1064~nm light sheet with the tight axis in the vertical direction compresses and holds the sample vertically.
%This should in principle help preventing the heating for excited state atoms due to anti-trapping during cooling and imaging, which usually occurs in near-detuned lattices.
%A beam tightly-focused in the vertical direction compresses the atom sample into a thin layer.
%The vertical confinement is provided by a tightly focused light sheet at 1064~nm, which compresses atoms into a thin 2D layer for our science experiment.

\begin{figure}
\centering
\includegraphics[width=85mm]{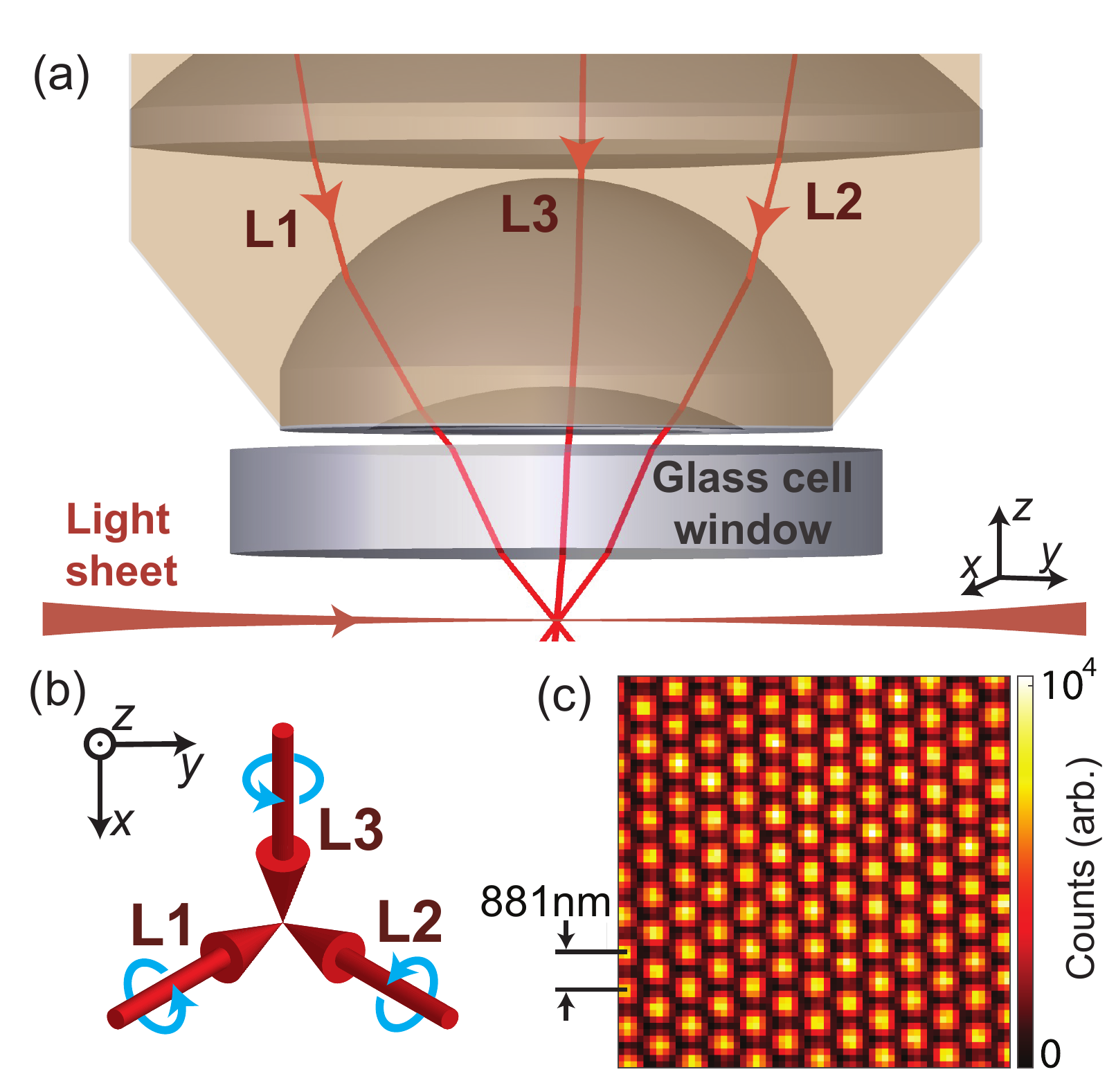}
\caption{Triangular optical lattice setup.
(a) We generate a 2D triangular lattice by intersecting three beams (L1, L2, L3) sent from the upper objective (light brown shaded area).
The light sheet propagating along $y$-axis provides a tight confinement in the vertical $z$-direction.
(b) The three lattice beams (red arrows) are circularly-polarized (blue arrows).
(c) The lattice intensity pattern directly measured on the diagnostic CCD.
The lattice constant is measured to be consistent with our prediction of 881~nm.
}
\label{figLattice}
\end{figure}

In the lattice setup, we split one laser beam (SolsTiS, M Squared Lasers) into three beams each with 0.25~W of power, and send them off-axis downwards through the upper objective (Fig.~\ref{figLattice}a).
They are aligned to be symmetric around the optical axis of the objective.
After the glass cell window, the three beams propagate inwards with an angle of $\theta = 45^\circ$ and intersect each other at a projected angle of 120$^\circ$ in the horizontal plane.
The interference of the three beams creates a triangular optical lattice with an expected lattice constant of $2\lambda/3\sin\theta = 881~\text{nm}$.
The $1/e^2$ beam radius at the crossing point is approximately 40~$\mu$m for all three beams.
%Atoms are trapped to the intensity maxima of the lattice.
After passing through the atoms and the lower objective, the three lattice beams are directed to a diagnostic CCD for real-time monitoring of the lattice potential on the atom plane.
%The interference pattern on the CCD is a magnified image of the 2D lattice at the objective focus, so from this we can infer lattice beam parameters and potential pattern at atoms location.

%The polarization of the lattice beams can be individually controlled.
%From our simulation, we found that the lattice pattern can change from triangular to hexagonal depending on the polarization configuration of the three beams.
%To achieve a deeper trap, we adopt the asymmetric linear polarization scheme from Ref.\cite{yang2021site}, which leads to a triangular trap geometry that is confirmed by the diagnostic CCD imaging (Fig.~4b and Fig.~4c).
To achieve the deepest lattice potential, we set all beams to be circularly polarized (Fig.~\ref{figLattice}b) to maximize the polarization overlapping between all three beams.
From the lattice beam parameters, we estimate a trap depth around $150~\mu\text{K}\approx 760~T_R$, where $T_R = 0.198~\mu\text{K}$ is the recoil temperature.
The lattice trap frequency is measured to be 75~kHz in the horizontal plane, consistent with our estimate.

The three beams share the same set of beam-shaping optics and propagate through almost-identical optical path lengths.
This setup contributes to the phase stability of the optical lattice and the pointing stability of beams at the crossing point.
%The relative phase drift among the three beams leads to translation of lattice sites.
%The optical path lengths for the three lattice beams are designed to be approximately equal so that it is insensitive to thermal expansion.
%The three beams also share the same set of beam-shaping optics to achieve better phase stability.
%This design also helps to maintain the beam size at the crossing point to be the same for all the three beams.

To confine atoms vertically, we apply a light sheet, which is a dipole trap formed by an elliptical Gaussian beam at 1064~nm with a power of 3~W (Mephisto MOPA, Coherent, Inc.) propagating along the $y$-axis (Fig.~\ref{figLattice}a).
%Check how to cite companies/equipment
The beam has a vertical waist of 3~$\mu$m and a horizontal waist of 70~$\mu$m, which provides a tight confinement in the vertical direction with an estimated trap frequency of 33~kHz, and a near-uniform intensity in the horizontal plane.
%Given the small vertical size and relatively high power
From measurements of the vertical beam size and laser power, we estimate the trap depth to be 3~mK.
The strong confinement ensures atoms are localized within the microscope's short depth of focus of $\approx$1~$\mu$m during the imaging.
%The dipole trap design does not suffer from multiple-layer loading as the conventional vertical lattice approach in many 2D lattice experiments, so the additional layer-selective procedure to remove the out-of-focus atoms can be bypassed.

\section{DMD Tweezer Array}
\label{secDMD}

\begin{figure}
\centering
\includegraphics[width=85mm]{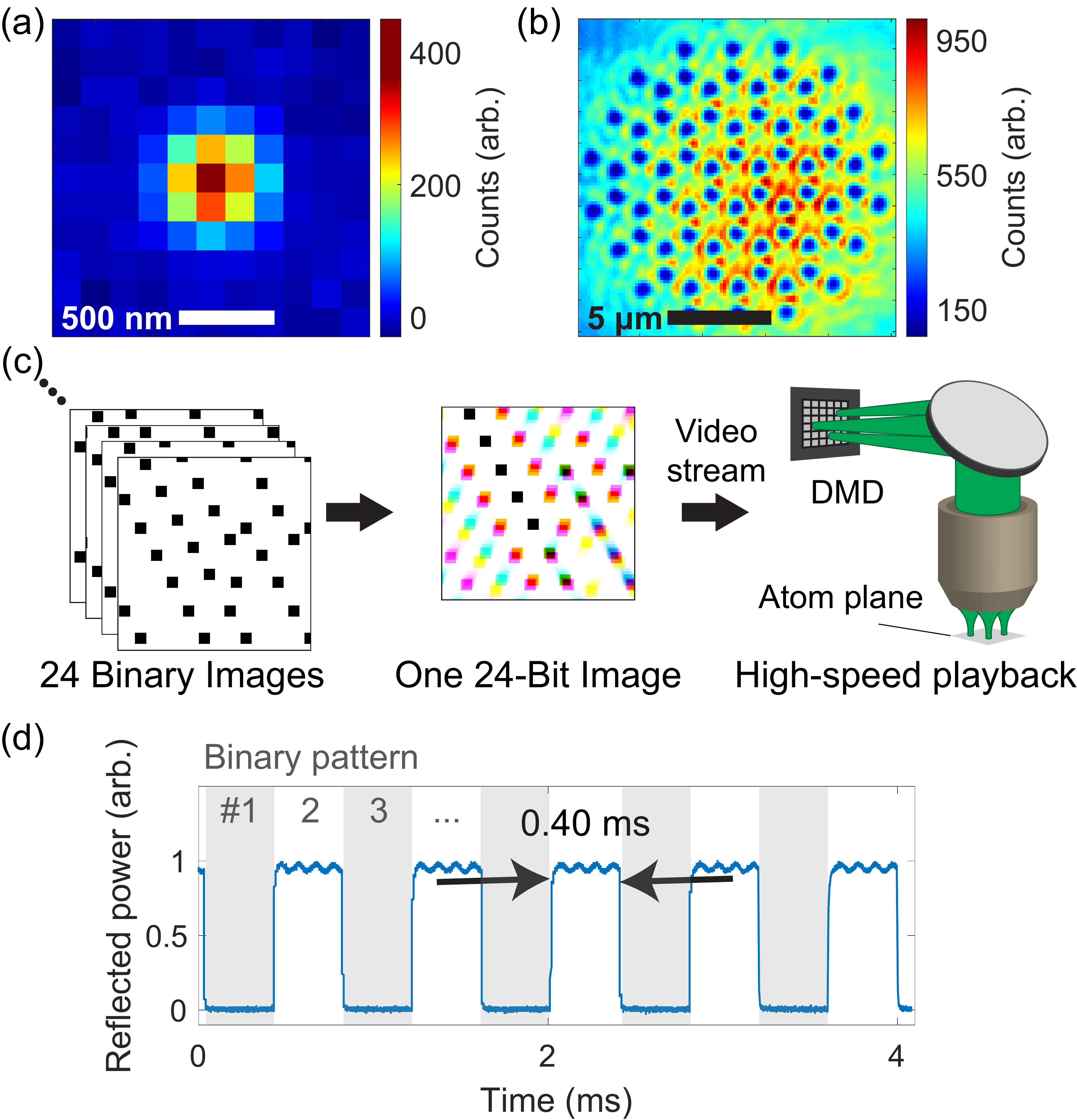}
\caption{Dynamical control of a tweezer array generated by a digital micromirror device (DMD).
(a) The spatial resolution of the DMD projection system can be estimated by taking an image of a DMD-generated point source.
The resolution based on the Rayleigh criterion is measured to be 450(25)~nm, which is near the diffraction limit.
(b) The DMD can generate arbitrary optical potentials, such as the 72-tweezer pattern shown here.
%adjacent sites are separated by 1.2 um
(c) Fast real-time pattern updating is realized by assembling every 24 binary patterns into a standard 24-bit color video-streaming signal, which results in a 2.52~kHz switching speed, see text.
(d) The photodiode signal shows the DMD operating at max speed. Here the DMD is switching between all-on and all-off binary patterns.
}
\label{figDMD}
\end{figure}

We rearrange atoms on the optical lattice using an array of mobile optical tweezers.
The tweezers are formed by reflecting a 532~nm laser (Sprout-Solo, Lighthouse Photonics) with a digital micromirror device (DMD), which is capable of projecting arbitrary intensity patterns on the atoms at high speeds\cite{ha2015roton,kim2016situ, stuart2018single}.

Our DMD (DLP4500, Texas Instruments) contains a 912$\times$1140 array of 7.64~$\mu$m square mirrors that can quickly flip between an on- and an off-state.
When a mirror is turned on, incident light is reflected into the upper microscope objective toward the atomic sample (Fig.~\ref{figQMS}b).
Light reflected from off-mirrors is sent to a beam dump.
Since 532~nm light is blue-detuned for the cesium $D_2$ transition, our tweezer array is formed by dark spots on a bright background (Fig.~\ref{figDMD}b).
Due to diffraction effects related to the periodic structure of the mirrors, 60$\%$ of the incident power is directed toward the atoms.

We place the DMD in the image plane of the upper microscope objective, so that the pattern of on-mirrors matches the intensity profile imposed on the sample.
%By using the image plane, we avoid switching many mirrors to apply local intensity changes, a requirement for spatial light modulators operating in the Fourier plane.
In our system, approximately 10 DMD mirrors correspond to one lattice spacing of 881~nm, providing sufficient spatial resolution for smooth tweezer motion.
%More precise number is 10.37 mirrors
Smooth motion is imperative to minimize the heating of atoms associated with discrete changes in the light intensity.

We characterize the resolution of the DMD projection system by imaging point-like patterns on the lower CCD at a wavelength of 532~nm.
%A typical pattern is a 3$\times$3 DMD-pixel square created using 532~nm light, which amounts to a 264~nm square on the microscope focal plane, an effective point source. <--put in caption
We fit the intensity pattern, which yields a near-diffraction-limited resolution of 450(25)~nm based on the Rayleigh criterion (Fig.~\ref{figDMD}a).
%The DMD can generate an arbitrary number of tweezers.
%An example pattern with 72 tweezers is shown in Fig.~5b.

%High-fidelity control of atoms requires that the optical tweezer array move at fast speeds and without significant delay time upon receiving instructions from the PC.
High-fidelity control of the tweezer array requires fast updating of the DMD pattern with minimal delay time.
To meet this requirement, we operate the DMD in a video-streaming mode where the computer treats the DMD as a display device (Fig.~\ref{figDMD}c).
To improve the streaming bandwidth, we package every 24 binary tweezer patterns into a single 24-bit color image.
The color images are then transmitted to the DMD at the maximum streaming rate of 105~Hz.
For each color image, the DMD plays the constituent 24 binary patterns in sequence, thus reaching a final update rate of the tweezer patterns of 24$\times$105~Hz = 2.52~kHz.
%whereupon the DMD will play the containing binary images in sequence at a speed of 24$\times$(105~Hz) = 2.52~kHz, where 105~Hz is the maximum refresh rate for streaming color images.
If we move the tweezers by translating their positions one micromirror at a time, then moving one atom to a neighboring site takes an estimated time of 4~ms, which is comparable to other tweezer rearrangement experiments~\cite{endres2016atom, PhysRevA.102.063107}.
%Lastly, the total delay time in streaming mode is measured to be only 30~ms, which is also relatively short.
Lastly, we measure the total delay time to transmit images from the computer to DMD to be only 30~ms, which is short compared to the atom lifetime.
%Fast dynamic control of tweezers is crucial in order successfully move atoms into the desired pattern.
%In order to realize real-time control of atoms based on their distribution in the lattice, we operate the DMD in a video-streaming mode.
%Expand below into ~three sentences
%In this mode, the DMD receives an HDMI video signal comprised of 24-bit images at a rate of 105~Hz from the control computer.
%Each 24-bit image contains 24 binary images that dictate the state of all mirrors on the DMD, which updates every 397~$\mu$s (24$\times$(105~Hz) = 2.52~kHz).

We measure the light reflected by the DMD on a fast photodiode in order to confirm the switching speed and playback quality.
By alternating the DMD mirrors between on and off states at max speed we observe the expected 2.52~kHz refresh rate (Fig.~\ref{figDMD}d).
We also confirm that long sequences of binary patterns are played in the correct order without pauses, skipping, or flickering~\cite{hueck2017note}.

%\begin{figure}[h]
\begin{figure}[b]
\centering
\includegraphics[width=85mm]{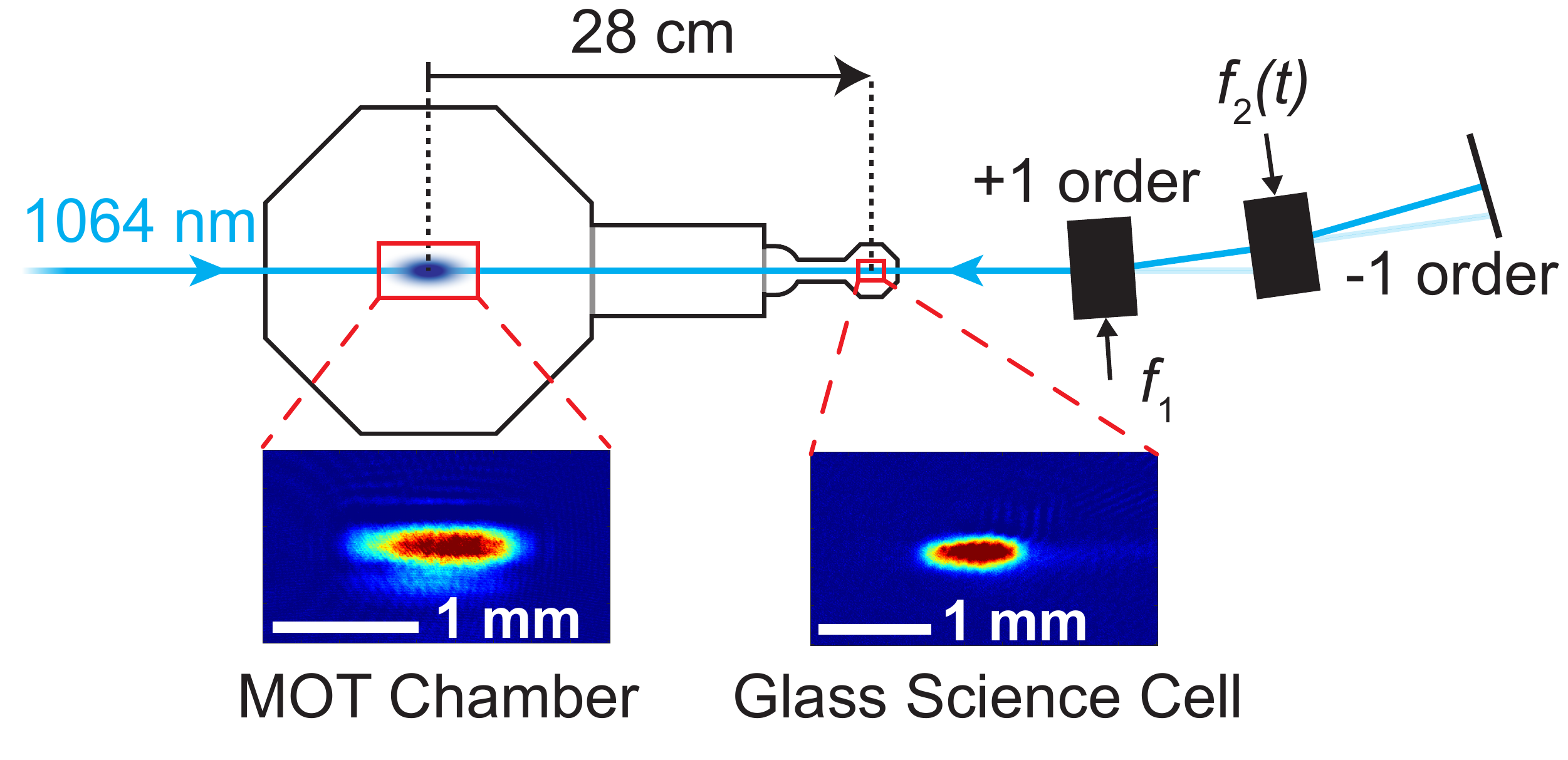}
\caption{Long distance transport of atoms based on a 1D optical lattice.
%Atoms are transported 28~cm from the stainless steel chamber to the glass science cell using a moving 1D optical lattice, which utilizes two AOM's to shift the frequency of the retro-reflected beam and establish time-dependent phase control of the lattice.
An incident beam at $\lambda_0=1064$~nm is retro-reflected, passing a pair of AOMs twice.
%A retro-reflected lattice beam passes a pair of AOMs twice and acquires a tunable frequency shift.
The first AOM is driven by a fixed frequency $f_1=80$~MHz while the second AOM operates at a tunable frequency $f_2(t) = 80\mathrm{ MHz}+\delta(t)$.
We take the $+1$ order from the first AOM and the $-1$ order from the second AOM, which results in the retro-reflected beam acquiring a frequency shift of $-2\delta(t)$.
We vary $\delta(t)$ along a smooth trajectory to translate the lattice sites 28~cm with speed $\delta(t)\lambda_0$.
Both the incident and retro-reflected beams have a beam waist of 300~$\mu$m near the center of transport, resulting in a long Rayleigh range and good trap uniformity to within 5$\%$ for the entire length.
The images show the initial atom cloud in the MOT chamber and the transported cloud in glass science cell.
}
\label{figTransport}
\end{figure}
%Lastly, we note that the reflected power shows no signs of ``flickering,'' an uncontrolled downward power spike that has been observed in various DMD models~\cite{hueck2017note}.

%Given the DMD-pixel size and our magnification, a tweezer can traverse one lattice constant in 4~ms.
%For a 10-by-10 atom sample, the tweezer re-arrangement is expected to be less than 50~ms, which is much faster than the atom sample lifetime.

\section{Long Distance Optical Transport of Atoms}
\label{secTran}

Our experiment starts with atoms pre-cooled in the MOT chamber before transferring them 28~cm to the glass cell (Fig.~\ref{figTransport}).
%Before reaching the microscope, atoms first undergo an initial cooling stage 28~cm away in a stainless steel vacuum chamber.
To transport the atoms, we load them into a 1D optical lattice and move the lattice sites by shifting the frequency of one of the lattice beams (see Refs.~\onlinecite{schmid2006long,klostermann2022fast}).

The 1D transport lattice is formed using light at wavelength $\lambda_0=1064$~nm with 40~W of power (Mephisto MOPA, Coherent, Inc.) and the frequency shift is realized using two acousto-optic modulators (AOMs) in a double-pass configuration, see Fig.~\ref{figTransport}.
%The first AOM is driven by a fixed frequency $\omega_1/(2\pi)=80$~MHz while the second AOM has a variable frequency $\omega_2(t)/(2\pi)$ near 80~MHz (Fig.~\ref{figTransport}).
%We take the $+1$ order from the first AOM and the $-1$ order from the second AOM, so that the lattice is stationary when both are driven by 80~MHz.
When a detuning $\delta(t) = f_2(t)-f_1$ is applied to the second AOM drive, the retro-reflected beam is frequency shifted by $-2\delta(t)$, which results in the lattice sites acquiring a velocity equal to $\delta(t)\lambda_0$.
%This all-optical scheme is insensitive to mechanical vibrations, and can be made much faster assuming one has a deep enough lattice.
This all-optical transport scheme does not require any moving parts, is insensitive to mechanical vibrations, and can proceed quickly with atoms in a deep lattice.

%\FloatBarrier
\begin{figure*}[t]
\centering
%\begin{center}
\includegraphics[width=150mm]{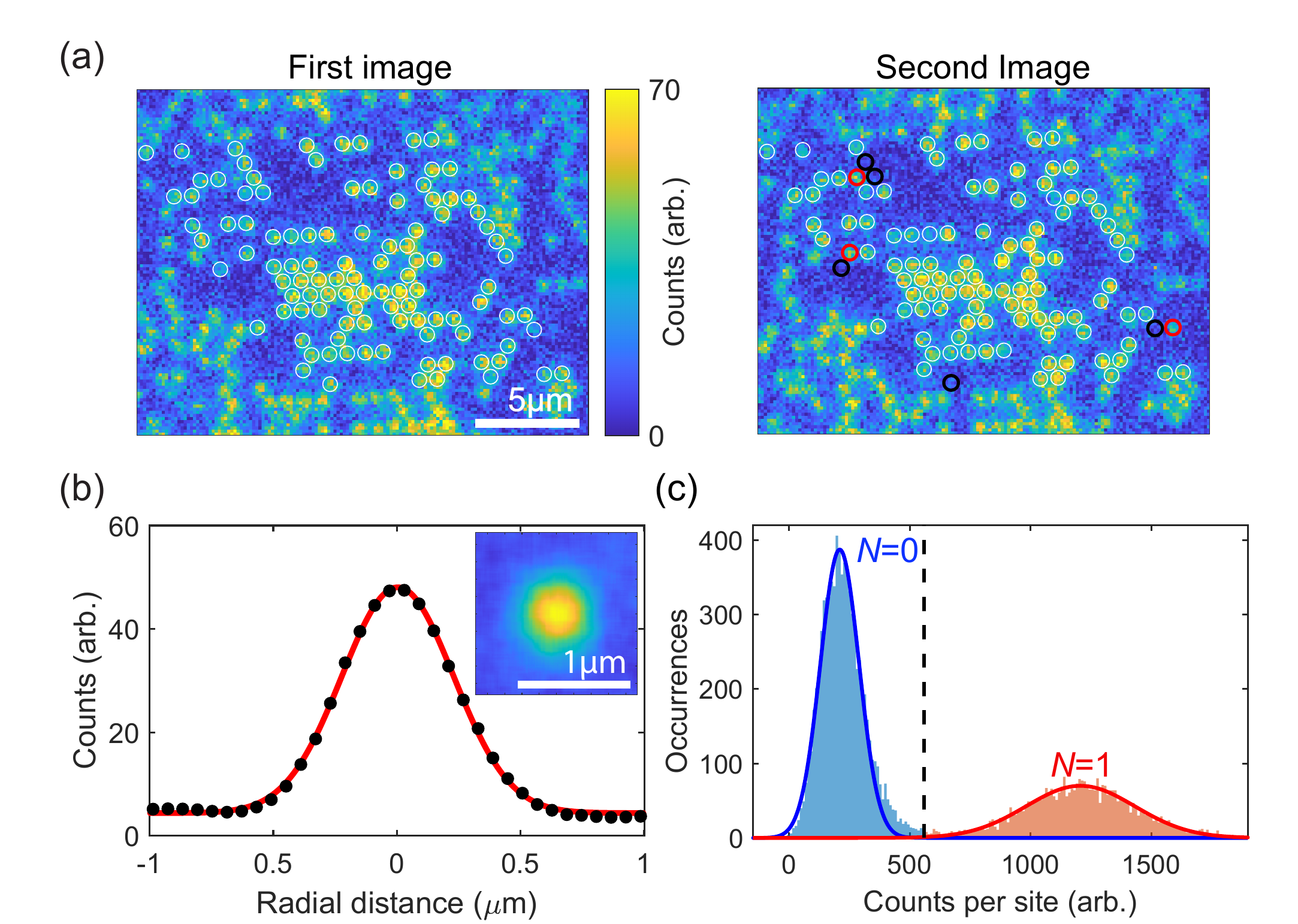}
\caption{Site-resolved fluorescence imaging.
(a) Typical raw fluorescence images taken 5~ms apart on the lower CCD with markers drawn in the central region to guide the eye. Each image has an exposure time of 200~
ms. 
Sites that were empty and became filled are marked red, and sites that were filled and became empty are marked black in the second image (right).
%In the second image (right), sites that became filled between the first and second image are marked with red circles, and sites that became empty are marked with black circles.
(b) Radial profile of a point source averaged over $180^\circ$ (black) and Gaussian fit (red). The width (based on the Rayleigh criterion) of 655(3)~nm indicates diffraction-limited performance. Inset: Averaged image of a point source generated using a sparsely-filled atom sample.
(c) Histogram of counts per site in the central region using data from 40 images shows a peak for unoccupied sites (blue, $N=0$) and a peak for occupied sites (red, $N=1$).
The blue and red lines are Gaussian fits, and the vertical dashed line defines the threshold for the determination of site occupancy.
}
\label{figSingleSite}
%\end{center}
\end{figure*}

In the experimental sequence, atoms in the MOT chamber are first cooled by degenerate Raman sideband cooling and loaded into the transport lattice (250~$\mu$K trap depth, 170~kHz trap frequency).
Immediately after cooling, $\delta(t)$ is smoothly increased to its final value of 600~kHz, reaching a max speed of 64~cm/s, after which it decreases back to zero so that the atom cloud gently comes to a stop at the center of the glass cell.
%Both the incident and retro-reflected beams have a beam waist of 300~$\mu$m near the center of transport, resulting in a long Rayleigh range and good trap uniformity to within 5$\%$ for the entire length.
We typically transport 20$\sim$30$\%$ of the initial atom number in 400~ms to the glass cell and end with 6$\times 10^5$ atoms.
%The retention rate is believed to be limited by the relatively slow 10~kHz update rate of the DDS, which causes atoms to experience a velocity kick every 100~$\mu$s.

%\FloatBarrier
\section{Cooling and site-resolved imaging}
\label{secImg}

Once atoms are transported into the glass cell, they are cooled into the 2D triangular lattice using dRSC.
Photons emitted during the cooling process are collected by the microscope objectives to produce site-resolved images of atoms on the CCDs.

%The dRSC scheme for cesium atoms closely follows Ref.~\cite{PhysRevLett.84.439}.
%The 2D-triangular-lattice beams as described in Section \ref{secLattice} provide both the trapping potential and the Raman coupling necessary to reduce the vibrational energy.
%Circularly polarized $\sigma^+$ optical pumping ($6S_{1/2}, F=3\rightarrow 6P_{3/2}, F'=2$) and repumping ($6S_{1/2}, F=4\rightarrow 6P_{3/2}, F'=4$) beams are applied in the same direction as the magnetic field to cool atoms toward the vibrational ground state.
%The magnetic field is in the horizontal plane bisecting the $x$- and $y$-axes, see Fig.~\ref{figLattice}b.
%The photons spontaneously emitted during the optical pumping process can be collected on the microscope objectives to obtain high-resolution images of the atoms.

%The atoms are prepared for imaging in the following way.
%Atoms transported from the MOT chamber are loaded into the 2D triangular lattice where an initial dRSC stage is performed.
After atom transport we perform an initial dRSC stage to cool the sample into the 2D triangular lattice.
The dRSC scheme closely follows Ref.~\onlinecite{PhysRevLett.84.439}.
In our system, the lattice beams as described in Section \ref{secLattice} provide both the trapping potential and the Raman coupling necessary to reduce the vibrational energy.
Circularly polarized $\sigma^+$ optical pumping ($6S_{1/2}, F=3\rightarrow 6P_{3/2}, F'=2$) and repumping ($6S_{1/2}, F=4\rightarrow 6P_{3/2}, F'=4$) beams are applied in the same direction as the magnetic field to complete the cooling cycle.
The magnetic field is in the horizontal plane bisecting the $x$- and $y$-axes as shown in Fig.~\ref{figLattice}b.
After cooling the sample for 20~ms, we reach a temperature of 5~$\mu$K in the lattice.

Following the initial cooling stage, the light sheet power is ramped up in 2~ms to compress the sample vertically into a thin layer.
The lattice potential is then ramped down to release excess atoms outside of the light sheet.
We then ramp the lattice power back up and perform a second dRSC stage.
During cooling, pairs of atoms are ejected via light-assisted collisions, resulting in each lattice site having only zero or one atoms.
Within 50~ms, the atoms are cooled near the 3D vibrational ground state after which we expose the CCDs to collect the photons. 
%We wait another 450~ms before exposing the CCDs to collect the photons.

A typical imaging exposure time is 200~ms, wherein an estimated 10$\%$ of the total emitted photons are collected on each of the upper and lower CCDs (iKon-M 934, Andor Technology Ltd) after accounting for the finite solid angle of the objectives and transmission losses.
We can take 8 or more images per experimental cycle in order to study the loss and tunneling behavior over time.
Fig.~\ref{figSingleSite}a shows an example of two consecutive exposures with a 5~ms hold time in between.

We analyze a central $25 \times 25$~$\mu$m region that contains approximately 200 lattice sites.
Using a sparsely-filled sample, we can overlap the signal from single atoms to obtain the point spread function, which yields an imaging resolution of 655(3)~nm based on the Rayleigh criterion.
Our result is in agreement with the diffraction limit given by $0.61\lambda_{D2}/ \mathrm{NA}=650$~nm, where $\lambda_{D2}=852$~nm is the wavelength of the emitted photons (Fig.~\ref{figSingleSite}b).

We extract the positions of the lattice sites by taking the Fourier transform of the atom images.
The sites are indicated with circular markers in Fig.~\ref{figSingleSite}a.
%The lattice vectors span an angle of $60(1)^\circ$, in good agreement with the expected triangular structure.
We determine the lattice constants to be 845~nm, 860~nm, and 874~nm, all within 5$\%$ of our expectations.
We attribute the discrepancies to imperfect lattice beam alignment.

We determine the atom occupancies by extracting the photon counts from each lattice site.
In order to obtain higher fidelity, we perform image deconvolution using a kernel method~\cite{parsons2016probing,xu2014inverse}.
With this method, a typical histogram of the photon counts is shown in Fig.~\ref{figSingleSite}c, and shows good separation between the distributions for sites with and without atoms.
By fitting both distributions to Gaussians, we can set a threshold value to evaluate the site occupancies.
The small Gaussian fits overlap suggests a high fidelity of 99$\%$ or an error rate of 1$\%$ from image reconstruction alone.

%To further evaluate the fidelity of single-site imaging, we take a series of up to eight images of the same atomic sample.
Our imaging allows us to study the particle dynamics over time by taking many images of the same atomic sample.
By comparing adjacent images, we can identify changes in the site occupations, see Fig.~\ref{figSingleSite}a for an example.
Disappearance of atoms in the later images is attributed to loss and hopping, while appearance of atoms in previously-unoccupied sites is likely due to hopping.
%Sites that are missing atoms are attributed to loss and hopping events, while sites where atoms appear are attributed to hopping (Fig.~\ref{figSingleSite}a).
%The loss is identified from the missing of the atoms in the later image compared to the previous image, and the hopping is identified from the fraction of atoms appearing in a previously-unoccupied site.
%From the measurement, we observe a loss fraction of 7.4$\%$ and a hopping fraction of 1.2$\%$.
The fidelity of atomic occupancies in two consecutive images is 91.4$\%$, limited by atom loss and tunneling.
%Want to: show images, details about fidelity (deconvolution, average PSF, loss/tunneling) 
%List of numbers to present:
%-Photons collected per atom (scattering rate)
%-Exposure time
%-Field of view, number of sites
%-Filling fraction (maybe)
%-Fidelity
%-PSF widths
%-Loss/tunneling rate
%-Lattice vector angles (to show triangularity)
%-Lattice site position stability (~many shots timescale)

%\section{Bitter Electromagnet}
%The low-inductance Bitter electromagnet design allows for high currents exceeding 100 A to be applied without significant heating due to the efficient water-cooling, which cools each copper layer in parallel.
%This avoids the usual temperature gradient found in a traditionally rolled coil where water flows along the wire and develops a warmer temperature when it reaches the end of the coil.
%The 3D-printed structure allows a more compact form-factor when compared to previous designs and allows for more space around the glass cell, which is a region dense with optical elements.

\section{Conclusion}
In summary, we have presented the design and characterization of the quantum matter synthesizer, a novel cold atom quantum simulation platform that combines the clean trapping potential of a 2D optical lattice with the site-resolved control of atoms using an optical tweezer array.
%Key demonstrations toward the final completion of the QMS were shown, including single-site atomic imaging and the high-speed, low-latency optical tweezer array.

%Numerous innovative design elements were incorporated, including the nanotextured anti-reflective glass cell, the formation of a lattice using off-axis beams, the all-optical atomic transport, and the dual-objective microscope system.
Numerous innovative design elements were incorporated, including the nanotextured anti-reflective glass cell, the formation of a triangular lattice through the microscope objective, the long-distance atom transport with a 1D moving lattice, and the dual-objective microscope system.
%The mechanical stability between the lattice and tweezers, a key metric of feasibility, has been shown to be smaller than 10~nm, a fraction of the on-site wavefunction extent.
By combining the lattice and tweezer beams into the objective, we reach an exquisite mechanical instability below 10~nm, a small fraction of the on-site wavefunction extent.
%A new type of dynamical tweezer array formed by a DMD has been tested, and it exhibits fast speeds and small delay times, meeting the requirements to perform quick rearrangement based on images obtained within an experimental shot.
We also present a new type of dynamical tweezer array formed by a DMD that exhibits fast streaming speeds and low latency.
Lastly, we demonstrate high-fidelity site-resolved imaging of Cs atoms in a 2D triangular lattice with relatively small exposure times of 200~ms.
Through these technical developments, the QMS is well-positioned to simulate many-body physics with precise state preparation and measurement fidelity.
%Fix sentence above plus add a nice closing sentence about many-body physics etc at single-particle level

\section*{Author Declarations}
The authors have no conflicts to disclose. 

\section*{Author Contributions}
J.T. designed and constructed the experiment, collected and analyzed data, and prepared the manuscript. M.Z. contributed to the construction of the experiment, collected and analyzed data, and prepared the manuscript. L.W. helped prepare the manuscript. C.C. supervised the project.

\section*{Acknowledgments}
We thank Gustaf Downs, Mickey McDonald, Kai-Xuan Yao, and Paloma Ocola for early work on the design and construction of the system. 
We also thank Mykhaylo Usatyuk and Samir Rajani for characterization of the DMD. 
This material is based upon work supported by the U.S. Department of Energy, Office of Science, Office of Basic Energy Sciences, under Award Number DE-SC0019216 and by the National Science Foundation Graduate Research Fellowship under Grant No. DGE 1746045.
%This material is based upon work supported by the National Science Foundation Graduate Research Fellowship under Grant No. DGE 1746045.

\section*{Data Availability}
The data that support the findings of this study are available from the corresponding author upon reasonable request.

%\clearpage
\bibliography{Refs.bib}

\end{document}